# CMS pixel telescope addition to T-980 bent crystal collimation experiment at the Tevatron

Ryan Rivera[a]*, Jerry Annala[a], Todd Johnson[a], Simon Kwan[a], Carl Lundberg[a], Dean Still[a], Alan Prosser[a], Lorenzo Uplegger[a], Jim Zagel[a], Viktoriya Zvodaya[a]

[a]*Fermi National Accelerator Laboratory, Batavia, IL 60510 USA*

**Abstract**

An enhancement to the T-980 bent crystal collimation experiment at the Tevatron has been completed. The enhancement was the installation of a pixel telescope inside the vacuum-sealed beam pipe of the Tevatron. The telescope is comprised of six CMS PSI46 pixel plaquettes, arranged as three stations of horizontal and vertical planes, with the CAPTAN system for data acquisition and control. The purpose of the pixel telescope is to measure beam profiles produced by bent crystals under various conditions. The telescope electronics inside the beam pipe initially were not adequately shielded from the image current of the passing beams. A new shielding approach was devised and installed, which resolved the problem. The noise issues encountered and the mitigating techniques are presented herein, as well as some preliminary results from the telescope.

* Ryan Rivera. Tel.: +1-630-840-3000.
 *E-mail address*: rrivera@fnal.gov.

## 1. Introduction

In the fall of 2010, a telescope consisting of six PSI46 pixel detectors [1] was installed in the Tevatron beam line at Fermi National Accelerator Laboratory in Batavia, Illinois for the T-980 Bent Crystal Collimation experiment. The ongoing experiment's focus is to study the effects of bent crystals for the application in collimation systems [2]. Halo particles can be channeled (CH) or volume reflected (VR) [3] within the crystal lattice deviating halo particles from the main beam by a given angle. The pixel telescope was added to image the CH or VR beam after it passes through the crystal.

A diagram of the T-980 experimental setup located in the E0 long straight section of the Tevatron ring is shown in Fig. 1. The telescope is installed just upstream of the E03 collimator.

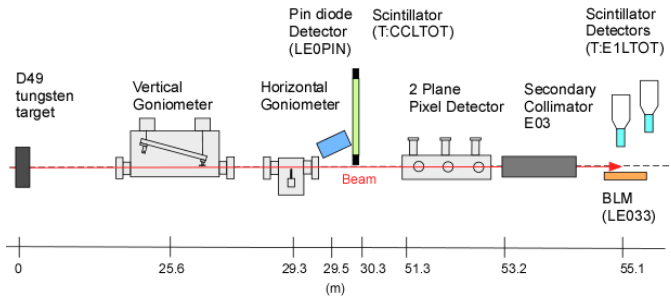

Fig. 1. Above is a sketch of the accelerator section where the experiment is installed. It is located at E0 in the Tevatron accelerator tunnel. The CMS pixel telescope addition is identified by the label, "2 Plane Pixel Detector." The collimating crystals are positioned in the beam using the vertical and horizontal goniometers.

The pixel detectors employed in the telescoping system are identical to those used in the forward pixel detectors of the Compact Muon Solenoid (CMS) experiment at the European Organization for Nuclear Research (CERN) in Geneva, Switzerland. Fig. 2 demonstrates the simplified process by which the pixel telescope will be placed to image the channeled beam from the bent crystal.

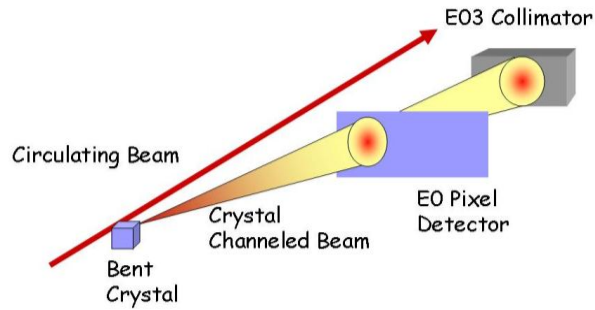

Fig. 2. Above is a depiction of the placement of the pixel telescope to intercept and image the channeled beam from the bent crystal.

## 2. The setup

The pixel telescope consists of six PSI46 pixel plaquettes installed inside a vacuum chamber. The design of the telescope is unique in that it is the first application of placing the PSI46 pixel plaquette in a high vacuum environment of approximately 1E-9 torr. In order to achieve this vacuum criteria, it is required that the pixel and vacuum chamber together be baked at a temperature of 100 C for five days. No degradation of the chip or wire-bonds was noticed due to the bake.

The Tevatron particle beam is comprised of both protons and antiprotons moving in opposite directions close to the center of the beam pipe. However, the proton beam will be the only beam that is affected by the bent crystal. Therefore, the pixel telescope which is normally placed out of the beam at the edge of the beam pipe must be moved into the beam and have the freedom to align to the channeled beam. The front and back of the vacuum vessel have motors for translating the pixels in the vertical and horizontal planes to approach the beam at various angles.

For clarity, consider the six pixel detectors as being arranged in three sets of two – with each set, or station, consisting of one horizontal detector and one vertical detector. From this perspective, the telescope has three stations: upstream, midstream, and downstream.

The locations of the six pixel detectors along the beam pipe are identified in Fig. 3.

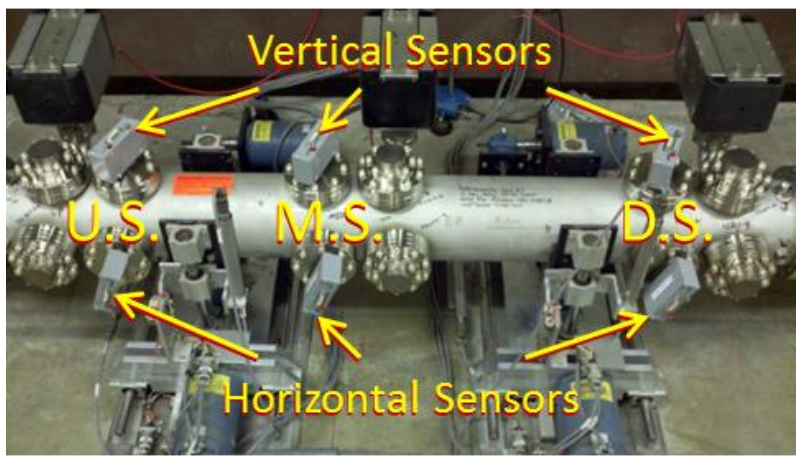

Fig. 3. The pixel vacuum chamber is shown here in situ in the Tevatron tunnel. The six CMS PSI46 pixel detectors are installed inside the vacuum chamber at the locations indicated by the arrows – there are three horizontal detectors and three vertical detectors. The abbreviations U.S., M.S., and D.S. identify the upstream, midstream, and downstream pixel stations respectively.

Each of the six detectors is actually a pixel plaquette comprised of two PSI46 readout chips sharing a single silicon sensor. The shared sensor yields a continuous sensitive area of 8.2 mm by 16.2 mm. The entire plaquette is 9.3 mm by 17 mm, and is mounted on and wire-bonded to a high density interconnect flex circuit laminated to 200 microns of silicon, which is 15.3 mm by 19.4 mm. This flex circuit is mounted to an aluminum mechanical support. A sketch of the pixel assembly is shown in Fig. 4.

Still inside the vacuum chamber, the flex circuit is wire-bonded to a 3 cm by 7 cm passive circuit board that connects to a 15 cm twisted-pair cable. All power lines, inputs, and outputs for a single plaquette are present on this twisted-pair cable, which leads to a DB25 feedthrough connector that exits the vacuum chamber. A snapshot of the assembly inside the vacuum chamber is shown in Fig. 5.

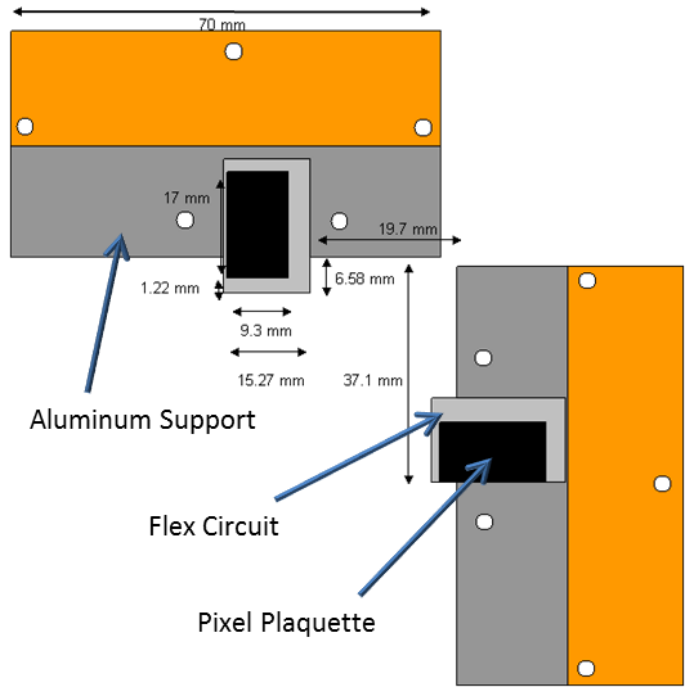

Fig. 4. Above is a sketch of the pixel assembly for one station. A station consists of two pixel plaquettes – one vertical and one horizontal. This drawing shows the dimensions and relative positions of the detector pair inside the vacuum chamber. The top layer is the pixel plaquette, followed by the flex circuit, and then the aluminum support.

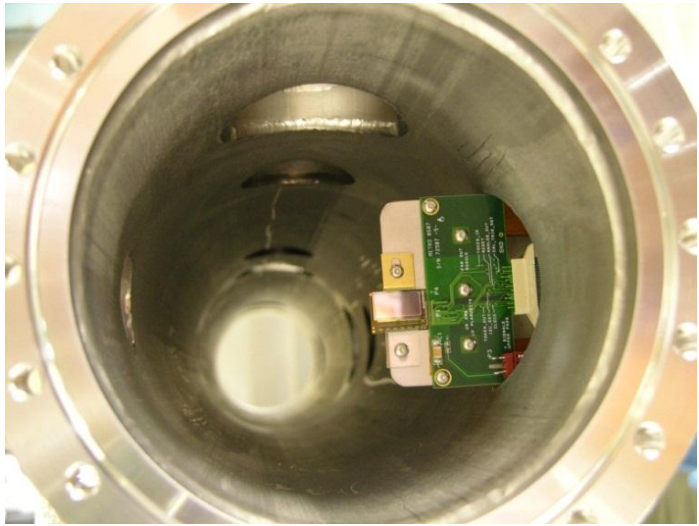

Fig. 5. This image is of a single pixel assembly within the vacuum chamber. The view is from upstream towards downstream through the center of the vacuum vessel.

Immediately outside the vacuum chamber and mounted to the feedthrough connector, is a printed circuit board designed by the Electronic Systems Engineering (ESE) department of the Computing Division (CD) at Fermilab. All signals for one plaquette are present on this board. The plaquette input signals run passively through the board to the vacuum through the DB25 connector. The output signals, from the plaquette, are amplified and buffered on this board using a radiation hard, amplifier layout with surface mount JFETs. All of the input and outputs are transmitted over a 3 m standard DVI cable between this amplifier board and the data acquisition system.

The pixel data acquisition and control is handled by the CAPTAN (Compact and Programmable daTa Acquisition Node) system [4, 5, 6]. Fig. 6 shows one of the CAPTAN electronics modules. Each one of these modules handles up to four plaquettes – two of the modules were deployed for this project.

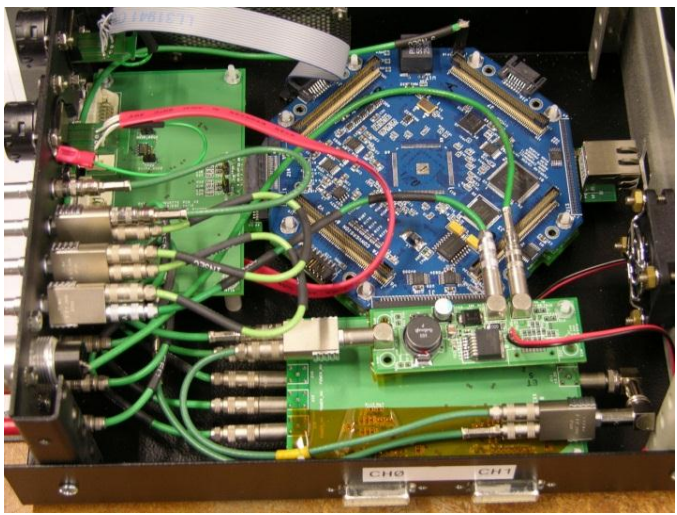

Fig. 6 Above is the CAPTAN data acquisition and control module used for the T-980 experiment to readout and control up to four pixel plaquettes.

The CAPTAN system is a flexible and scalable hardware system with communication and processing achieved using the Xilinx Virtex-4 FPGA Family and gigabit Ethernet components. For this experiment, the CAPTAN modules sit in the tunnel in the overhead cable rack, and provide signaling and power to the pixels through a DVI cable.

Each of the two CAPTAN systems is directly connected to a computer at the surface through a single commercial 50 meter CAT5 Ethernet cable. There is also one adjustable power supply feed for each CAPTAN from the surface (~14V @ 2.5A) and an additional CAT5 Ethernet cable repurposed to carry JTAG signals for configware modifications to the FPGA. The CAPTAN software suite runs on the computer at the surface that is on the other end of the 50 meter Ethernet cables. The software can also be controlled remotely through other computers on the network.

### 3. The problem

After installation, the electronics functioned as designed and communication between the computer and the pixels was established. Everything seemed to go as planned until the Tevatron beam was turned on. At which point, reliable transmission of the control signals to the pixels could no longer be achieved, and the analog output lines from the plaquettes exhibited large noise spikes.

It became clear that this noise could be attributed to the particle bunch pattern of the Tevatron. This noise pattern can be seen in Fig. 7.

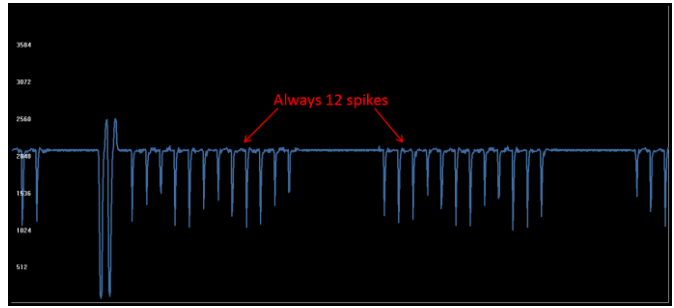

Fig. 7. This figure shows the noisy analog output from one plaquette as digitized by the CAPTAN system at approximately 50 MHz. The timing of the noise spikes is consistent with the Tevatron particle bunch pattern.

The noise on the analog output of a plaquette is a problem because this output is where all of the pixel data is encoded. The pixel data is comprised of the row and column information that indicates the physical position of a particle passing through the sensor. The row and column information is integral to assembling a meaningful image of the channeled beam that has passed through the pixels. And so, this analog signal must be interpreted precisely for the pixels to have any utility for T-980.

The standard interpretation algorithm failed to decipher row and column information reliably in the presence of the aforementioned noise spikes.

## 4. The solution

Two approaches were contemplated to mitigate the noise issue: a software solution and a hardware solution. The software modification was attempted first for two reasons. One, there is less risk of irreversible damage to the setup. And two, a software solution could be completed faster than designing, fabricating, and installing hardware, which also involved waiting for Tevatron downtime to access the tunnel.

The software solution employed was a software-based filter. The goal of the filter was to reduce the effects of the spikes while maintaining the signal integrity of the pixel output signal. The results of the filter are shown in Fig. 8.

The software-based noise filtering solution attempted to identify and soften the spikes using a median filter. This was not straightforward because the noise spikes occur near the same frequency as the level changes that must be decoded. Additionally, since the pixel system was not synchronized with the Tevatron accelerator clock, the relative position of the spikes to the data was not fixed. So the distortions to the data caused by the spikes were unpredictable across readout cycles.

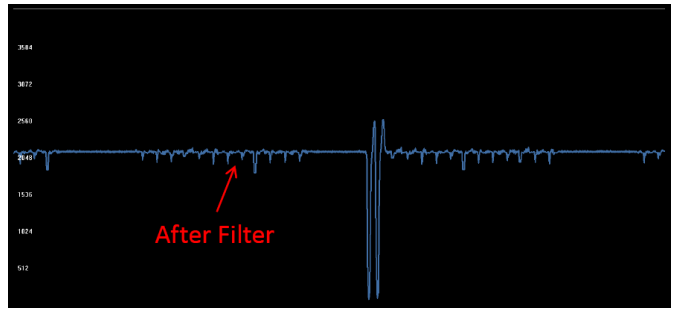

Fig. 8. This figure shows the analog output from one plaquette after the software-based filter approach was applied. It can be seen that the spikes were reduced substantially when compared to Fig. 7.

After several iterations, the software filter managed to significantly improve data reconstruction; however reconstruction was still not perfect. Also, the software-based solution did nothing to address the noise on the control commands sent to the chips. Without dependable control lines, it was not possible to have confidence in the configuration of the pixels.

It was decided that making the hardware modification was worth the risk to reduce the noise on the pixel inputs and outputs. And so the second approach was undertaken. The second approach consisted of changing the route of the twisted pair cable and applying more robust shielding to the signals inside the vacuum chamber. The twisted pair cable originally ran from the pixel port hole, along the inside of the vacuum chamber for approximately six inches before exiting an auxiliary port hole. The modification extended the pixel port hole radially to allow the pixels and the cable to fit within the vacuum chamber at a single point.

The resulting hardware configuration inside the vacuum chamber is shown in Fig. 9.

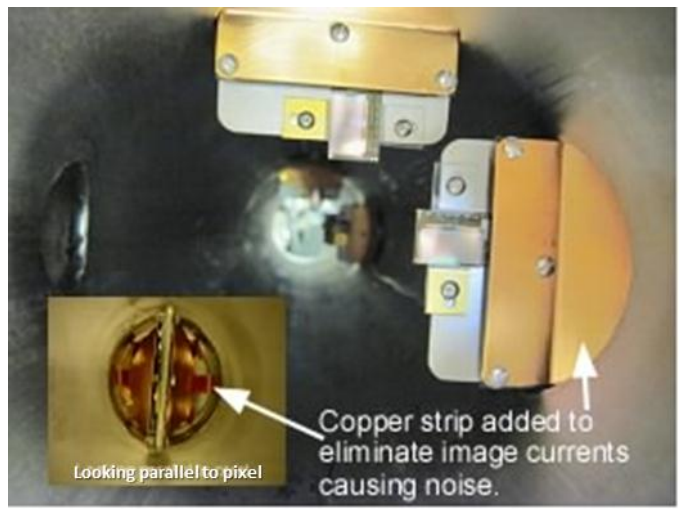

Fig. 9. This image is of a pixel station within the vacuum chamber with copper shielding applied to provide a path for the beam image current and reduce noise. The view is from upstream towards downstream through the center of the vessel.

Copper plates were devised that mounted over the pixel assembly. The copper plates cover all of the pixel circuitry except for the sensor and make electrical contact with the inside of the vacuum canister.

Shielding in this manner provides a path around the pixel detectors for the image current of the passing protons and antiprotons that make up the Tevatron particle beam – protecting the pixel circuitry from playing this role. The final result was a clean signal as displayed in Fig. 10.

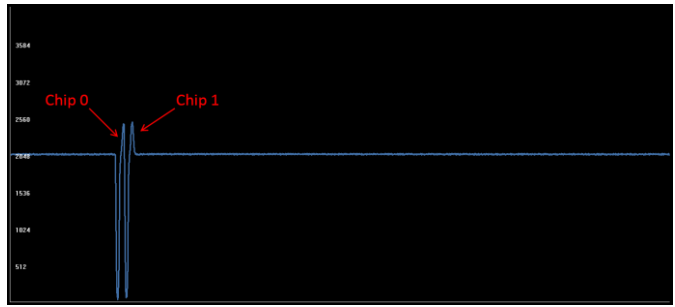

Fig. 10. This figure shows the analog output from one plaquette after the hardware approach to noise mitigation was applied. It can be seen that the noise spikes are completely removed when compared to Fig. 7 and Fig. 8.

## 5. Conclusion

On May 18, 2011, channeled beam from the bent crystal was successfully imaged on a single plaquette in the Fermilab Tevatron. A one second accumulation of pixel data with beam channeled from the bent crystal can be seen in Fig. 11.

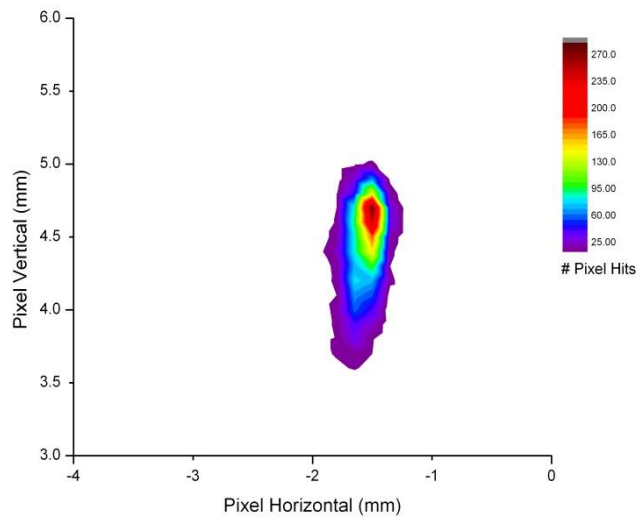

Fig. 11. This figure shows the successful imaging of the channeled beam in the pixels. The bright spot is the channeled beam which is displaced from the main beam core by approximately 4.5 mm with dimensions of approximately 1 mm by 3 mm.

Unfortunately, only the downstream horizontal pixel array could be used to gather meaningful data. During the process of adding the noise mitigating copper plates and re-vacuum certifying the vessel, four of the six pixel detectors were damaged. This only left one working pixel array in each plane. This made it impossible to use the full tracking capability of the telescope, which requires at least two detectors in the same plane. Because of the time constraints imposed by the run schedule of the Tevatron, the damaged pixel arrays were left in place when the vacuum vessel was re-installed. A full autopsy of the broken pixel arrays will have to wait until the end of the Tevatron run on October 1, 2011.